\documentclass[sn-mathphys-num]{sn-jnl}

\usepackage{graphicx}%
\usepackage{multirow}%
\usepackage{amsmath,amssymb,amsfonts}%
\usepackage{amsthm}%
\usepackage{natbib}%
\usepackage{upgreek}%
\usepackage{textgreek}%
\usepackage{mathrsfs}%
\usepackage[title]{appendix}%
\usepackage{xcolor}%
\usepackage{textcomp}%
\usepackage{manyfoot}%
\usepackage{booktabs}%
\usepackage{algorithm}%
\usepackage{algorithmicx}%
\usepackage{algpseudocode}%
\usepackage{listings}%
\usepackage{gensymb}%
\usepackage[colorinlistoftodos]{todonotes}%

\begin{document}

\title[Article Title]{2D-3D Deformable Image Registration of Histology Slide and Micro-CT with ML-based Initialization}


\author*[1,2]{\fnm{Junan} \sur{Chen}}\email{chen@imfusion.com}

\author[1]{\fnm{Matteo} \sur{Ronchetti}}

\author[3]{\fnm{Verena} \sur{Stehl}}

\author[1]{\fnm{Van} \sur{Nguyen}}

\author[3]{\fnm{Muhannad Al} \sur{Kallaa}}

\author[4]{\fnm{Mahesh Thalwaththe} \sur{Gedara}} 

\author[4]{\fnm{Claudia} \sur{L\"olkes}}

\author[4]{\fnm{Stefan} \sur{Moser}} 

\author[3]{\fnm{Maximilian} \sur{Seidl}} 

\author[1]{\fnm{Matthias} \sur{Wieczorek}} 

\affil[1]{\orgname{ImFusion GmbH}, \orgaddress{\street{Agnes-Pockels-Bogen 1}, \city{Munich}, \postcode{80992}, \country{Germany}}}

\affil[2]{\orgdiv{Department of Physics}, \orgname{Technical University of Munich}, \orgaddress{\street{James-Franck-Str. 1}, \city{Garching}, \postcode{85748}, \country{Germany}}}

\affil[3]{\orgdiv{Institute of Pathology}, \orgname{Heinrich Heine University and University Hospital of D\"usseldorf}, \orgaddress{\street{Mooren Str. 5}, \city{D\"usseldorf}, \postcode{40225}, \country{Germany}}}

\affil[4]{\orgname{Fraunhofer Institute for High-Speed Dynamics, Ernst-Mach-Institut}, \orgaddress{\street{Ernst Zermelo Str. 4}, \city{Freiburg}, \postcode{79104}, \country{Germany}}}

\abstract{Recent developments in the registration of histology and micro-computed tomography (\textmu CT) have broadened the perspective of pathological applications such as virtual histology based on \textmu CT. This topic remains challenging because of the low image quality of soft tissue CT. Additionally, soft tissue samples usually deform during the histology slide preparation, making it difficult to correlate the structures between histology slide and \textmu CT. In this work, we propose a novel 2D-3D multi-modal deformable image registration method. The method uses a machine learning (ML) based initialization followed by the registration. The registration is finalized by an analytical out-of-plane deformation refinement. The method is evaluated on datasets acquired from tonsil and tumor tissues. \textmu CTs of both phase-contrast and conventional absorption modalities are investigated. The registration results from the proposed method are compared with those from intensity- and keypoint-based methods. The comparison is conducted using both visual and fiducial-based evaluations. The proposed method demonstrates superior performance compared to the other two methods.}


\maketitle

\section{Introduction}\label{sec1}

Histology and CT are two imaging modalities commonly integrated in clinical applications. Histological imaging is widely used to examine tissue samples to study their morphology or composition on a microscopic scale, which is a key procedure in clinical pathology and histopathology. Various staining methods are used to enhance the contrast for different tissue components such as lymphatic tissue, connective tissue, and muscle. In Germany, around 40 million tissues are analyzed by 1800 pathologists each year for cancer diagnosis \cite{pathologists}. However, given the relatively long turnaround time (exceeding 3 days for the conventional, established standard procedure in pathology laboratories \cite{TAT1} \cite{TAT2}) and the large number of tissue examination per pathologist each year, there is an increasing need from the clinics to improve the efficiency of the current cancer diagnosis pipeline.  

Focusing on improving the time cycle of the diagnosis pipeline, various studies on virtual histology using CT have been carried out \cite{Hieber2016} \cite{Lee2022}. CT is another favored clinical imaging modality given its non-destructive feature and rapid acquisition of 3D data. Recent developments of the CT imaging setup \cite{microCT} enable the resolution under sub-micrometer scale using absorption and phase-contrast modalities. The novel CT imaging methods reveal the possibility of resolving micro-structures that are usually visible in histology images but invisible in conventional clinical CTs. With the histology image and the matching \textmu CT slice, one can transfer annotations by pathologists from histology to \textmu CT, which is a key step to realize virtual histology. However, the process of searching the matching \textmu CT slice is non-trivial. The histology slides are acquired following the conventional sample preparation and sectioning procedure \cite{histology}. The histological sectioning is not always parallel to the surface of the sample block, a small angle ($\pm$ 5 degree) is manually selected by the pathologist while performing the cut \cite{cuttingangle}. During the sectioning process, the 2D slices undergo in- and out-of-plane deformation due to the blade motion \cite{SCHULZ201117}. To successfully enhance this classical, established laboratory workflow with CT-generated 3D images, this deformation must be unambiguously and computationally reversible, to create a clear link between coordinates in the \textmu CT volume and actual locations in the physical tissue sample. The core challenge in matching the images involves implementing a 2D-3D multi-modal deformable image registration algorithm.

Multi-modal deformable image registration is of great importance in medical image processing \cite{medicalimageprocessing}. Studies focusing on the 2D histology to 3D \textmu CT registration have been conducted in recent years \cite{currentapproaches}. To overcome the limit when evaluating the similarity between images of different modalities, an approach using ML-based modality translation with cascaded plane selection and 2D-3D deformable registration has been investigated \cite{Leroy}. Moreover, a method using speeded-up robust features (SURF) and scale-invariant feature transform (SIFT), which were originally invented for optical image matching, has also been developed \cite{feature1}\cite{feature2}. This method was developed with bone CTs. However, \textmu CTs of soft tissues have a lower contrast and a higher noise level in comparison to that of bone CTs. Moreover, a greater deformation exists between the \textmu CT and histology slide of the soft tissues. SURF and SIFT are sensitive to nonrigid transformation and to the noise of the image \cite{Lowe2004}. The performance of SIFT therefore drops drastically with soft tissue images. A method targeting on clinical soft tissue is not yet well established. The reproducibility of a fully ML-integrated registration framework is not always guaranteed when using images of tumor tissues as input, since the morphology varies dramatically between different tissues. Therefore, we propose a novel registration algorithm combining an ML-based initialization and an analytical optimization of the sampling plane.

The proposed 2D-3D deformable registration framework consists of plane initialization, plane refinement, and out-of-plane deformation optimization. Intensity- and keypoint-based initialization methods were previously investigated \cite{imxp}\cite{xnpig}. The first experiment exploits a relatively small capture range. The convergence highly depends on the quality of the initial guess. Recently, an ML-based differentiable similarity measure approximation (DISA) \cite{DISA} was introduced. Aiming at efficiently enlarging the capture range in the multi-modal image registration problem, this method has proven its effectiveness. In this work, we integrate a DISA-based initialization in our registration, which exhibits a better convergence in comparison with the approaches using intensity- and keypoint-based initialization.

\section{Methods}\label{sec2}
\begin{figure}[ht!]
\centering\includegraphics[width=\textwidth]{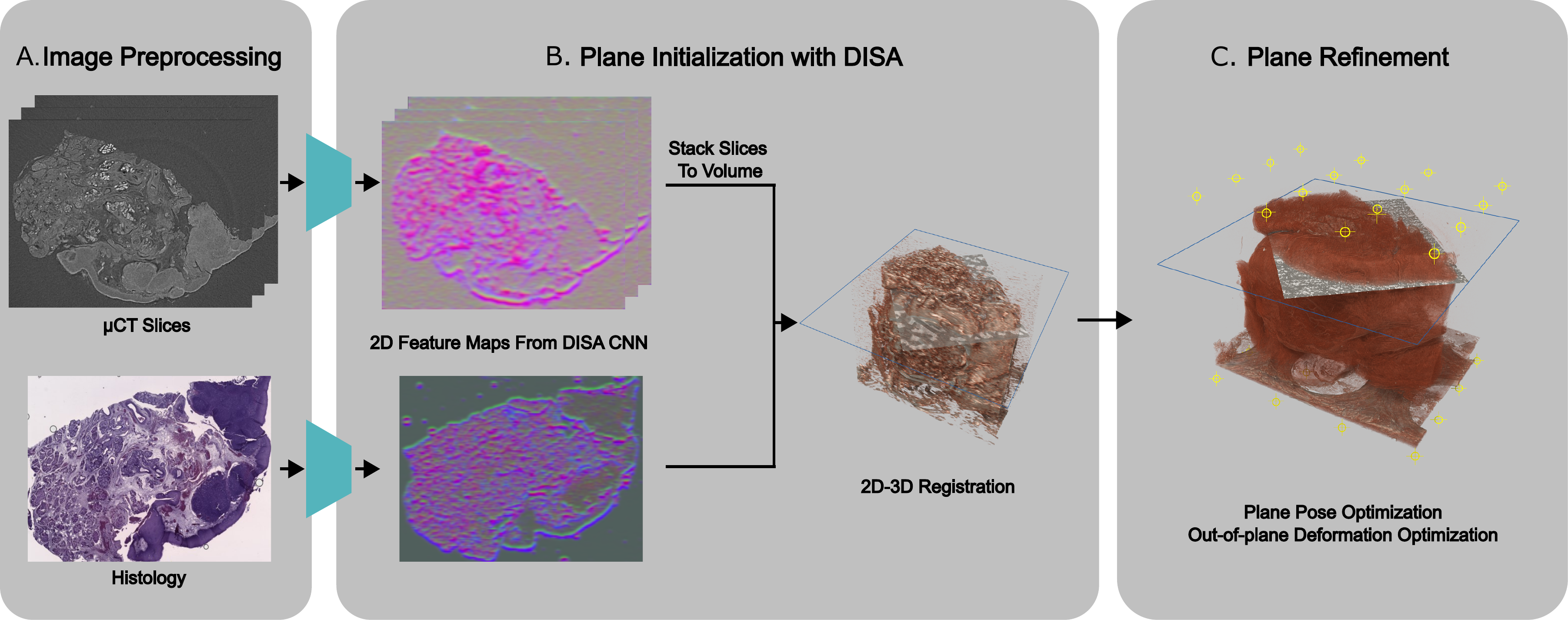}
\caption{The demonstration of the workflow of the proposed registration algorithm. (A) Image preprocessing with grayscale conversion and percentile normalization. (B) Initialization of the plane pose using 2D-3D registration on the feature maps of CT and histology. The original feature maps contain 16 channels, but only the first 3 channels are extracted for better visualization. (C) Sampling plane refinement by optimizing over the plane pose parameters and the out-of-plane deformation. Control point pairs are displayed as yellow markers above and under the CT volume.}
\label{fig:workflow}
\end{figure}
The 2D-3D histology to \textmu CT registration algorithm is equivalent to searching an optimal sampling plane which extracts a slice from the CT volume that matches the 2D histology slide. The algorithm consists of the following parts: plane initialization, optimization of plane pose parameters, and the optimization of the out-of-plane deformation. The workflow of the algorithm is demonstrated in Figure \ref{fig:workflow}. 

In the pre-processing phase (Figure \ref{fig:workflow}A), the histology image is converted to grayscale by computing the mean of all three channels. Additionally, percentile normalization using a range of 1\% to 99\% is applied to both the grayscale histology image and the \textmu CT. 

\subsection{Initialization}
We previously implemented an intensity-based initialization approach \cite{imxp}. This approach iteratively extracts 2D CT slices along one axis to and runs a 2D-2D image registration with the histology slide. The optimal slice is then defined as the one that returns the highest similarity with the histology slide. Here, the local normalized cross-correlation (LNCC) is used as the similarity measure. Besides, we have also implemented a keypoint-based initialization method \cite{xnpig}. Following the approach originally invented by Chicherova \textit{et al.}\cite{feature2}, 2D slices are iteratively extracted from the CT volume. Keypoints from the CT slice and the histology slide are then detected using the SIFT feature detector. The detected features are later matched and used for the plane fitting to search for the initial plane. The results from these approaches are compared later with that of the proposed method.

In this work, a novel initialization approach is introduced. The initialization is replaced with a high-performance image registration using DISA-$\text{LC}^2$ similarity measure. $\text{LC}^2$ similarity measure has been developed for multi-modal registration \cite{LC2}, which exhibits large capture range in the multi-modal problem. Given two images(namely source image and target image), the $\text{LC}^2$ similarity is computed patch-wise. $\text{LC}^2$ defines the similarity between the images by evaluating how well a target patch can be reproduced by a linear combination of the source patch and the gradient magnitude of the source patch. However, the $\text{LC}^2$ metric uses pixel-wise least-squares fitting method, which is therefore non-differentiable and computationally expensive \cite{LC22}. Thus, it is not ideal to use the primitive $\text{LC}^2$ metric for initialization purposes. We, therefore, train a 2D convolutional neural network (CNN) to approximate the $\text{LC}^2$ similarity by the dot product of the 2D feature maps of the histology slide and the CT slice. 

Extracting the CT slices from the volume as the input of the CNN, the output 2D CT feature maps are later stacked to a volume, whereas the 2D histology feature map is obtained with the 2D histology image as input. The thickness of the histology feature map is set to the estimated thickness (1.5 \textmu m) of the histology slide. 2D-3D image registration with DISA-$\text{LC}^2$ is performed on the two feature maps, whereby the histology feature map is set as the moving image, as shown in Figure \ref{fig:workflow}B. The resulting 3D transformation matrix of the histology feature map is then used as the pose of the initial plane, which serves as the initial guess for the later plane refinement optimization.
\subsection{Plane refinement}
\label{sec:planerefinement}
The optimization problems in the plane fine-tuning phase can be defined as follows:
\begin{equation} \label{Optimization}
\underset{\mathbf{T} \in \mathbb{R}^3}{\text{arg max}} ~~ s(I_{histo}, I_{ct}(\mathbf{T})),
\end{equation}
\begin{equation}
\underset{\mathbf{D} \in \mathbb{R}^N}{\text{arg max}} ~~ s(I_{histo}, I_{ct}(\mathbf{T}_{opt}, \mathbf{D})).
\end{equation}
Assuming extracting the CT slices along the z-axis, $\mathbf{T} \in \mathbb{R}^3$ denotes the pose parameters that transform the sampling plane in 3D space (translation on z-, rotations on x-and y-axis). $\mathbf{T}_{opt}$ refers to the optimal pose parameters returned from Eq. \eqref{Optimization}. We use a free-form deformation model based on B-spline \cite{FFD}. Two control points are assigned above and below the sampling plane as a pair. These pairs of control points are distributed in a grid pattern across the entire plane, as shown in Figure \ref{fig:workflow}C. By introducing displacements to the control point pairs, out-of-plane deformation of the plane is achieved. $\mathbf{D} \in \mathbb{R}^N$ denotes the z-displacements of the control point pairs and $N$ is the number of the grid points. $I_{histo}$ is the histology slide, $I_{ct}(\mathbf{T})$ and $I_{ct}(\mathbf{T}_{opt}, \mathbf{D})$ refer to the CT slices extracted from the sampling plane defined by the corresponding plane pose parameters and the out-of-plane deformation. The similarity between the 2D images after performing a 2D-2D in-plane deformable image registration is denoted as $s(\cdot ,\cdot)$.

\section{Experiments}\label{sec3}
The samples used in this work were human tonsil and tumor tissues prepared by the University Hospital of D\"usseldorf. The samples were embedded in paraffin blocks, sent for CT scans, and then slices were extracted from the blocks to prepare for the histology slides. Both absorption and phase-contrast \textmu CT were acquired and tested for the performance of the proposed algorithm. At the time of writing this paper, a limited amount (4 phase-contrast \textmu CTs and 2 absorption \textmu CTs) of \textmu CT-histology image pairs were available for training and analysis. The sample orientation is fixed for all CTs, with the block surface being roughly perpendicular to the z-axis. Slices were extracted via the established standard slicing procedure in a pathology laboratory. Despite the actual cutting direction not being perfectly parallel to the surface, it intrinsically provides initial information to decide which axis to extract the CT slices for later processing.

Following the architecture design proposed by Ronchetti \textit{et al.}\cite{DISA}, our model consists of 2D convolutional layers with LeakyReLU activation function \cite{LeakyReLU} and residual blocks \cite{7780459}, and employs BlurPool \cite{blurpool} for downsampling, leading to a combined striding factor of 4.

\subsection{Datasets and training}\label{sec3.1}
Two phase-contrast \textmu CTs and the two corresponding histology slides are used for training the CNN. We use a total of 640 phase-contrast \textmu CT slices with a spacing of 2.6 \textmu m acquired at a commercial phase-contrast \textmu CT setup Exciscope \cite{exciscope}. Two histology slides prepared by the University Hospital of D\"usseldorf are used. Since the approach does not require a ground truth matching between the CT and histology image, 640 CT-histology 2D image pairs are then generated by reusing the same two histology slides. We resample the CT volumes and the histology images to a spacing of 10.4 \textmu m and normalize the pixel values with a mean of 0 and standard variation of 1. Following the original sampling and training strategy for 3D images \cite{DISA}, we adapt the approaches for 2D images. The source-target patch sampling procedure is repeated 10000 times on each CT-histology image pairs, which results in a total of 40000 pairs of patches. Training converges to an average $\text{L}_2$ error of 0.0090 and validation converges to 0.0095. The output of the proposed CNN is a 16-channel image.
\subsection{Optimization}\label{sec3.3}
For the 2D-3D registration of the feature maps, we use the BFGS \cite{bfgs} optimizer. We equidistantly select 10 depths below the surface of the tissue block and repeat rigid and affine registration starting from the corresponding depth. The range of each registration trial is set to $\pm 10\degree$ and $\pm 400$\textmu m. The best result among the 10 trials is then selected to continue with the plane refinement later.

To improve the efficiency, we crop out the sample-free area in the CT volumes and the histology slides and resample the images to a same spacing as mentioned in section \ref{sec3.1}, which is approximately a downsampling factor of 4 for the CT volumes. We use derivative-free optimizer BOBYQA \cite{BOBYQA} from NLopt library \cite{NLopt} for the plane refinement optimizations. For the plane pose optimization, the iteration is aborted if the change of parameters falls below $1\times 10^{-4}$. For the out-of-plane deformation optimization, a grid of $4 \times 4$ for the control point pairs is used. To increase the search space, we repeat the optimization 5 times with random initial guess for the control point displacements and terminate each after 80 iterations. The best result among the 5 optimizations is then selected. For the 2D-2D deformable registration wrapped in the optimization step, a B-spline free-form deformation with a $4 \times 4$ control point grid is used. We use LNCC as similarity measure and BOBYQA as optimizer.

\section{Results and discussion}\label{sec4}
\begin{figure}[!ht]
\centering\includegraphics[width=\textwidth]{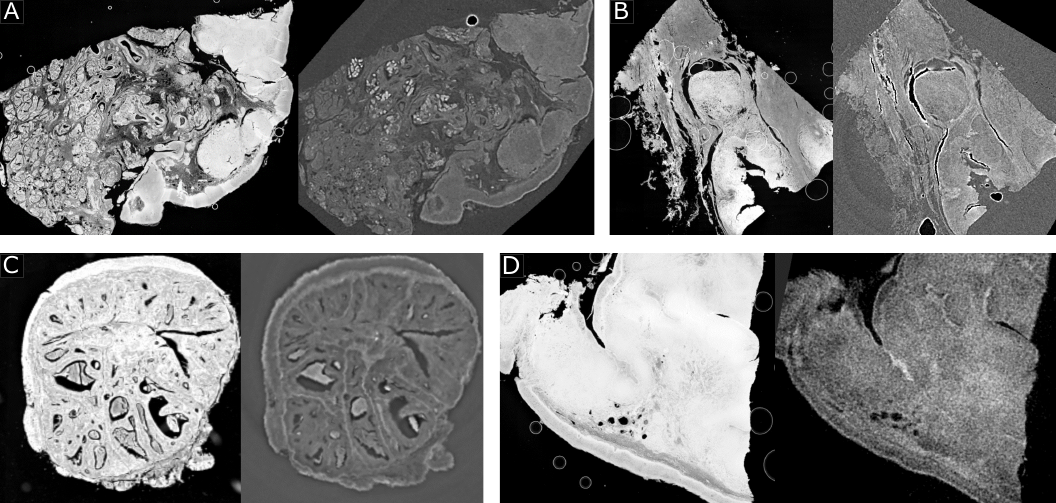}
\caption{Preprocessed histology slides and the registered CT slices of different tissues and imaging modalities. (A) and (B) Tonsil tissues and phase-contrast CTs. (C) Tumor tissue and phase-contrast CT. (D) Tonsil tissue and absorption CT.}
\label{fig:result_images}
\end{figure}

\begin{figure}[ht!]
\centering\includegraphics[width=\textwidth]{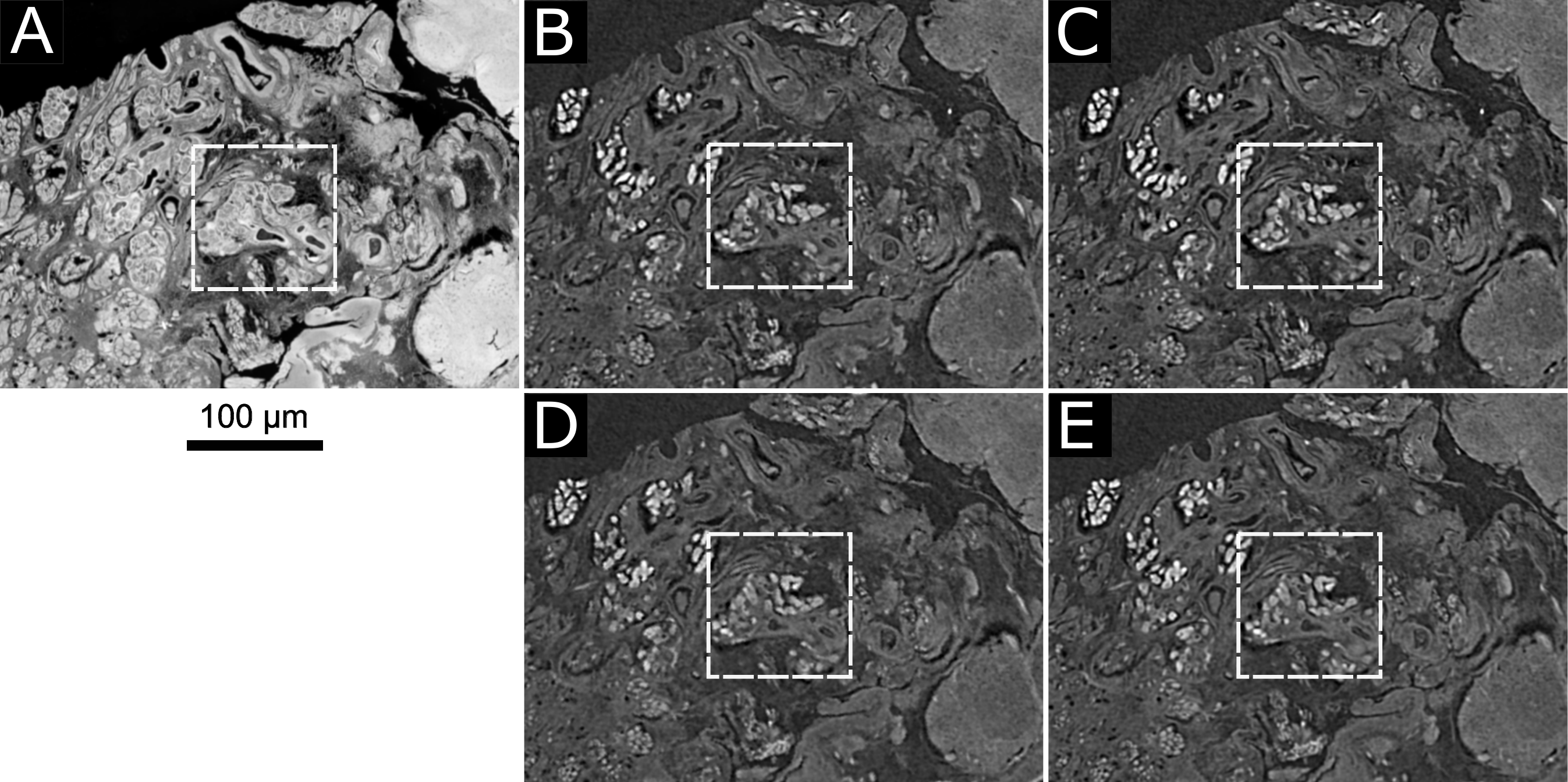}
\caption{Comparison of registration results from different initialization approaches. (A) Preprocessed histology image. (B) Registration using intensity-based initialization. (C) Registration using keypoint-based initialization. (D) Registration using DISA initialization. (E) Registration using manual initialization.}
\label{fig:result_comparison}
\end{figure}

\begin{figure}[ht!]
\centering\includegraphics[width=\textwidth]{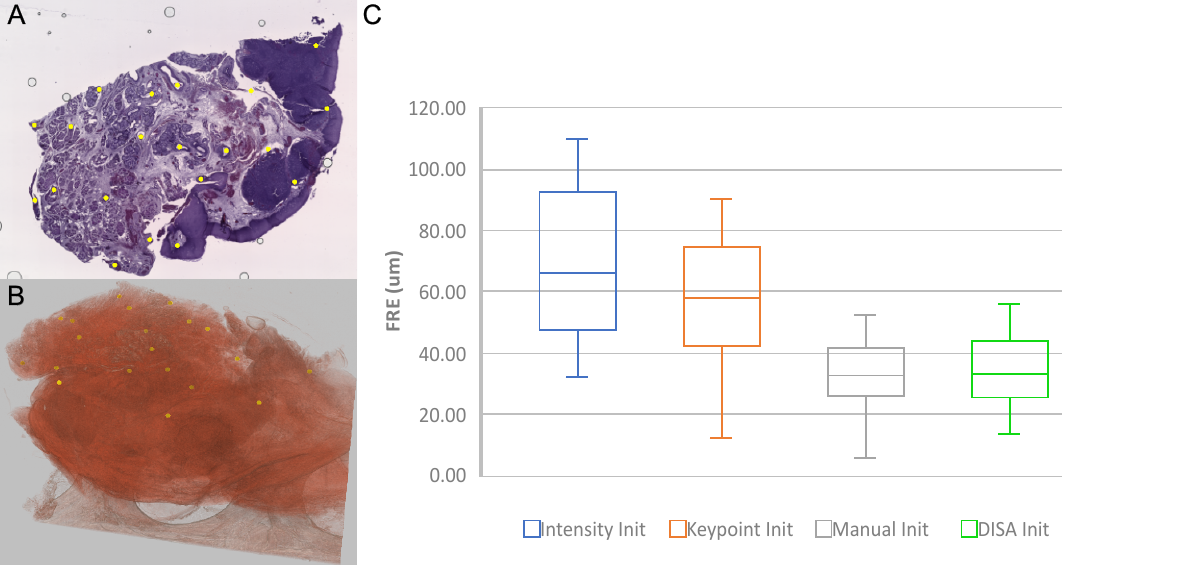}
\caption{The selected 20 fiducial pairs on the sample shown in Figure \ref{fig:result_images}A and the plot of the computed FRE values. (A) 20 manually annotated fiducials on the 2D histology. (B) Corresponding 20 fiducials on the CT volume. (C) FRE values of the registration results from four different initialization approaches}
\label{fig:fre}
\end{figure}

The proposed algorithm is applied to the rest of the datasets that are excluded from the model training, including 3 phase-contrast \textmu CT-histology pairs and 1 absorption \textmu CT-histology pair. These datasets exhibit different anatomical structures and image qualities than the training data. An evaluation of this approach is conducted based on the comparison of with the previously developed intensity- and keypoint-based methods. Similarity metrics LNCC and $\text{LC}^2$ are computed on the registered CT slice and histology slide. Fiducial pairs are selected manually from the CT and histology. After performing the registration, fiducial registration error (FRE) is computed as the mean distances between the fiducial marker pairs. 

On a CT volume with a dimension of $650 \times 580 \times 490$ and a resampled histology slide with a dimension of $780 \times 620$, the runtime of the initialization is 2 seconds for the DISA approach, which is significantly more efficient than the intensity-based initialization (30 seconds) and keypoint-based initialization (22 seconds).

The registered images from 4 datasets are shown in Figure \ref{fig:result_images}. The CTs in Figures \ref{fig:result_images}A and B are the histology slides and phase-contrast \textmu CTs of tonsil tissues acquired at Exciscope. The initial spacing of the CT volumes is 2.6 \textmu m. Figure \ref{fig:result_images}C shows the histology slide and the phase-contrast CT of tumor tissue acquired with the TOMCAT beamline at the synchrotron setup at Paul Scherrer Institut \cite{psi}, with a spacing of 1.6 \textmu m. An absorption \textmu CT acquired at MITOS GmbH \cite{mitos} and the corresponding histology slide are shown in Figure \ref{fig:result_images}D.  It shows visually good registration on various tissue types (tonsil, Figure \ref{fig:result_images}A, B, D, and tumor tissues, Figure \ref{fig:result_images}C) and imaging modalities (absorption, Figure \ref{fig:result_images}D, and phase-contrast CT, Figure \ref{fig:result_images}A, B, C), which indicates the proposed method has the potential to be further applied to different tissues and imaging modalities without retraining the model. All the registration results demonstrate visually good registration of the proposed approach, especially on the macro structures.

Detailed comparison on the registration results from different approaches are shown in Figure \ref{fig:result_comparison}. Note that these approaches only use different initialization methods, but they use the same method for plane refinement optimization, as mentioned in section \ref{sec:planerefinement}. Figure \ref{fig:result_comparison}B and C are the results from the previously developed approaches using intensity-based and keypoint-based initialization. In these results, the registration of specific micro-structures within the marked dashed square is not as robust as that of those larger structures outside of the square. Figure \ref{fig:result_comparison}D is the registered CT slice acquired from the proposed approach, which exhibits a better registration on the micro-structures than the previous results. As a reference, we also manually locate an initial plane that visually matches the histology slide the best and start the plane refinement optimization from this manual initial guess. The corresponding result is presented in Figure \ref{fig:result_comparison}E. To quantitatively compare the different approaches, we present the similarity values using LNCC and $\text{LC}^2$ metrics between the histology slide and the registered CT slice from the sample displayed in Figure \ref{fig:result_images}A. As shown in Table \ref{tab:compare_sm}, DISA initialization approach shows a higher value in both LNCC and $\text{LC}^2$ in comparison with the intensity- and keypoint-based approaches, which align well with the visual comparison on the images. 

\begin{table}[ht!]
  \centering
  \label{tab:compare_sm}
  \begin{tabular}{c c c c c}
    \toprule
     & Intensity-based Init & Keypoint-based Init & Manual Init & DISA Init \\
    \midrule
    LNCC & 0.37 & 0.38 & 0.47 & 0.45 \\
    $\text{LC}^2$ & 0.22 & 0.23 & 0.29 & 0.26 \\
    \bottomrule
  \end{tabular}
  \caption{Similarity values between histology slide and registered CT slices from the sample displayed in Figure \ref{fig:result_images}A.}
\end{table}

A total of 20 fiducial pairs on the histology and CT are annotated, as shown in Figure \ref{fig:fre}A and \ref{fig:fre}B. The corresponding FRE values from different approaches are computed after the registration, which are shown in the plot in Figure \ref{fig:fre}C. The FRE from the proposed approach exhibits a better registration result in comparison with the previous two approaches. It is at a similar level with the FRE from the reference manual approach.

\section{Conclusions and Outlook}\label{sec5}
In this work, a novel algorithm for 2D-3D multi-modal image registration is developed. With the proposed initialization method, the results demonstrate that the convergence of the optimization outperforms the intensity- and keypoint-based methods, with a significantly improved runtime of initialization. Using the unsupervised training, one can apply the DISA CNN model directly on images acquired from different tissues and modalities without the need for retraining the model. The accuracy of the registration results is comparable to that of the manual results. It shows the proposed method can be potentially applied for generating ground-truth matching image pairs, which are essential to training networks for modality translation and modality synthesis.

Despite the orientation of the sample surface being utilized as initial information for the algorithm, the workflow can be further automated by integrating an intensity-based surface detection algorithm. The out-of-plane deformation is realized by applying the deformation model on the CT volume. The CT slices are then extracted from the deformed volume via a flat sampling plane, which is mathematically equivalent to deforming the sampling plane directly. The runtime of the plane refinement optimization could be drastically improved in future by deforming the sampling plane instead of the CT volume. Moreover, the integration of DISA in the plane refinement phase will be accomplished. In the actual histology-CT image pairs, one can observe more complicate deformation and discontinuity, such as folding and fracturing. Some local micro-structures are still not perfectly registered. The deformation at the micro-structures can be modelled better with a denser control point grid, however, the optimization will become more expensive with an increased degree of freedom. Future study is needed to improve the complexity of the deformation model while maintaining a decent computational efficiency. Furthermore, the discontinuity in the actual deformation should also be considered. 

\section{Acknowledgements}\label{sec6}
This study was partially supported by the German Ministry for Education and Research (BMBF), grant number 13GW0571A, “HORUS”. Collaborating institutions: ImFusion GmbH, Munich; DORNER Health IT Solutions, Muellheim; Fraunhofer EMI, Efringen-Kirchen; Institute for Surgical Pathology, Medical Center-University of Freiburg; Institute of Pathology, Heinrich Heine University and University Hospital of D\"usseldorf

\bibliography{bibliography}


\begin{thebibliography}{30}
\ifx \bisbn   \undefined \def \bisbn  #1{ISBN #1}\fi
\ifx \binits  \undefined \def \binits#1{#1}\fi
\ifx \bauthor  \undefined \def \bauthor#1{#1}\fi
\ifx \batitle  \undefined \def \batitle#1{#1}\fi
\ifx \bjtitle  \undefined \def \bjtitle#1{#1}\fi
\ifx \bvolume  \undefined \def \bvolume#1{\textbf{#1}}\fi
\ifx \byear  \undefined \def \byear#1{#1}\fi
\ifx \bissue  \undefined \def \bissue#1{#1}\fi
\ifx \bfpage  \undefined \def \bfpage#1{#1}\fi
\ifx \blpage  \undefined \def \blpage #1{#1}\fi
\ifx \burl  \undefined \def \burl#1{\textsf{#1}}\fi
\ifx \doiurl  \undefined \def \doiurl#1{\url{https://doi.org/#1}}\fi
\ifx \betal  \undefined \def \betal{\textit{et al.}}\fi
\ifx \binstitute  \undefined \def \binstitute#1{#1}\fi
\ifx \binstitutionaled  \undefined \def \binstitutionaled#1{#1}\fi
\ifx \bctitle  \undefined \def \bctitle#1{#1}\fi
\ifx \beditor  \undefined \def \beditor#1{#1}\fi
\ifx \bpublisher  \undefined \def \bpublisher#1{#1}\fi
\ifx \bbtitle  \undefined \def \bbtitle#1{#1}\fi
\ifx \bedition  \undefined \def \bedition#1{#1}\fi
\ifx \bseriesno  \undefined \def \bseriesno#1{#1}\fi
\ifx \blocation  \undefined \def \blocation#1{#1}\fi
\ifx \bsertitle  \undefined \def \bsertitle#1{#1}\fi
\ifx \bsnm \undefined \def \bsnm#1{#1}\fi
\ifx \bsuffix \undefined \def \bsuffix#1{#1}\fi
\ifx \bparticle \undefined \def \bparticle#1{#1}\fi
\ifx \barticle \undefined \def \barticle#1{#1}\fi
\bibcommenthead
\ifx \bconfdate \undefined \def \bconfdate #1{#1}\fi
\ifx \botherref \undefined \def \botherref #1{#1}\fi
\ifx \url \undefined \def \url#1{\textsf{#1}}\fi
\ifx \bchapter \undefined \def \bchapter#1{#1}\fi
\ifx \bbook \undefined \def \bbook#1{#1}\fi
\ifx \bcomment \undefined \def \bcomment#1{#1}\fi
\ifx \oauthor \undefined \def \oauthor#1{#1}\fi
\ifx \citeauthoryear \undefined \def \citeauthoryear#1{#1}\fi
\ifx \endbibitem  \undefined \def \endbibitem {}\fi
\ifx \bconflocation  \undefined \def \bconflocation#1{#1}\fi
\ifx \arxivurl  \undefined \def \arxivurl#1{\textsf{#1}}\fi
\csname PreBibitemsHook\endcsname

\bibitem[\protect\citeauthoryear{Bruno et~al.}{2021}]{pathologists}
\begin{barticle}
\bauthor{\bsnm{Bruno}, \binits{M.}},
\bauthor{\bsnm{Laszló}, \binits{F.}},
\bauthor{\bsnm{Ralf}, \binits{H.}},
\bauthor{\bsnm{Svenja}, \binits{B.}},
\bauthor{\bsnm{Tina}, \binits{S.}}:
\batitle{Number of pathologists in germany: comparison with european countries,
  usa, and canada}.
\bjtitle{Virchows Arch}
\bvolume{478},
\bfpage{335}--\blpage{341}
(\byear{2021})
\end{barticle}
\endbibitem

\bibitem[\protect\citeauthoryear{Dias and Barrett}{2023}]{TAT1}
\begin{barticle}
\bauthor{\bsnm{Dias}, \binits{M.M.}},
\bauthor{\bsnm{Barrett}, \binits{A.W.}}:
\batitle{Comparison of histopathological turnaround times for mandibulectomies,
  glossectomies, and incisional biopsies of the tongue}.
\bjtitle{British Journal of Oral and Maxillofacial Surgery}
\bvolume{61}(\bissue{2}),
\bfpage{131}--\blpage{135}
(\byear{2023})
\end{barticle}
\endbibitem

\bibitem[\protect\citeauthoryear{Sharma et~al.}{}]{TAT2}
\begin{botherref}
\oauthor{\bsnm{Sharma}, \binits{A.}},
\oauthor{\bsnm{Nishadham}, \binits{V.}},
\oauthor{\bsnm{Gupta}, \binits{P.}},
\oauthor{\bsnm{Gupta}, \binits{G.}},
\oauthor{\bsnm{Sharma}, \binits{D.}},
\oauthor{\bsnm{Goel}, \binits{S.}},
\oauthor{\bsnm{Pasricha}, \binits{S.}},
\oauthor{\bsnm{Kamboj}, \binits{M.}},
\oauthor{\bsnm{Mehta}, \binits{A.}}:
Evaluation of turnaround times of diagnostic biopsies: A metric of quality in
  surgical pathology.
International Journal of Surgical Pathology
\textbf{0}(0)
\end{botherref}
\endbibitem

\bibitem[\protect\citeauthoryear{Hieber et~al.}{2016}]{Hieber2016}
\begin{barticle}
\bauthor{\bsnm{Hieber}, \binits{S.E.}},
\bauthor{\bsnm{Bikis}, \binits{C.}},
\bauthor{\bsnm{Khimchenko}, \binits{A.}},
\bauthor{\bsnm{Schweighauser}, \binits{G.}},
\bauthor{\bsnm{Hench}, \binits{J.}},
\bauthor{\bsnm{Chicherova}, \binits{N.}},
\bauthor{\bsnm{Schulz}, \binits{G.}},
\bauthor{\bsnm{M{\"u}ller}, \binits{B.}}:
\batitle{Tomographic brain imaging with nucleolar detail and automatic cell
  counting}.
\bjtitle{Scientific Reports}
\bvolume{6}(\bissue{1}),
\bfpage{32156}
(\byear{2016})
\end{barticle}
\endbibitem

\bibitem[\protect\citeauthoryear{Lee et~al.}{2022}]{Lee2022}
\begin{barticle}
\bauthor{\bsnm{Lee}, \binits{J.Y.}},
\bauthor{\bsnm{Mack}, \binits{A.F.}},
\bauthor{\bsnm{Shiozawa}, \binits{T.}},
\bauthor{\bsnm{Longo}, \binits{R.}},
\bauthor{\bsnm{Tromba}, \binits{G.}},
\bauthor{\bsnm{Scheffler}, \binits{K.}},
\bauthor{\bsnm{Hagberg}, \binits{G.E.}}:
\batitle{Microvascular imaging of the unstained human superior colliculus using
  synchrotron-radiation phase-contrast microtomography}.
\bjtitle{Scientific Reports}
\bvolume{12}(\bissue{1}),
\bfpage{9238}
(\byear{2022})
\end{barticle}
\endbibitem

\bibitem[\protect\citeauthoryear{Pacil{\`e} and Tromba}{2018}]{microCT}
\begin{bbook}
\bauthor{\bsnm{Pacil{\`e}}, \binits{S.}},
\bauthor{\bsnm{Tromba}, \binits{G.}}:
In: \beditor{\bsnm{Giuliani}, \binits{A.}},
\beditor{\bsnm{Cedola}, \binits{A.}} (eds.)
\bbtitle{Introduction to X-Ray Micro-tomography},
pp. \bfpage{19}--\blpage{39}.
\bpublisher{Springer},
\blocation{Cham}
(\byear{2018})
\end{bbook}
\endbibitem

\bibitem[\protect\citeauthoryear{Funkhouser}{2018}]{histology}
\begin{bchapter}
\bauthor{\bsnm{Funkhouser}, \binits{W.K.}}:
\bctitle{Chapter 11 - pathology: The clinical description of human disease}.
In: \beditor{\bsnm{Coleman}, \binits{W.B.}},
\beditor{\bsnm{Tsongalis}, \binits{G.J.}} (eds.)
\bbtitle{Molecular Pathology},
\bedition{Second edition} edn.,
pp. \bfpage{217}--\blpage{229}.
\bpublisher{Academic Press},
\blocation{Rockville, MD, USA}
(\byear{2018})
\end{bchapter}
\endbibitem

\bibitem[\protect\citeauthoryear{Dey}{2018}]{cuttingangle}
\begin{bbook}
\bauthor{\bsnm{Dey}, \binits{P.}}:
\bbtitle{Tissue Microtomy: Principle and Procedure},
pp. \bfpage{41}--\blpage{50}.
\bpublisher{Springer},
\blocation{Singapore}
(\byear{2018})
\end{bbook}
\endbibitem

\bibitem[\protect\citeauthoryear{Schulz et~al.}{2011}]{SCHULZ201117}
\begin{barticle}
\bauthor{\bsnm{Schulz}, \binits{G.}},
\bauthor{\bsnm{Crooijmans}, \binits{H.J.A.}},
\bauthor{\bsnm{Germann}, \binits{M.}},
\bauthor{\bsnm{Scheffler}, \binits{K.}},
\bauthor{\bsnm{Müller-Gerbl}, \binits{M.}},
\bauthor{\bsnm{Müller}, \binits{B.}}:
\batitle{Three-dimensional strain fields in human brain resulting from formalin
  fixation}.
\bjtitle{Journal of Neuroscience Methods}
\bvolume{202}(\bissue{1}),
\bfpage{17}--\blpage{27}
(\byear{2011})
\end{barticle}
\endbibitem

\bibitem[\protect\citeauthoryear{Sotiras et~al.}{2013}]{medicalimageprocessing}
\begin{barticle}
\bauthor{\bsnm{Sotiras}, \binits{A.}},
\bauthor{\bsnm{Davatzikos}, \binits{C.}},
\bauthor{\bsnm{Paragios}, \binits{N.}}:
\batitle{Deformable medical image registration: A survey}.
\bjtitle{IEEE Transactions on Medical Imaging}
\bvolume{32}(\bissue{7}),
\bfpage{1153}--\blpage{1190}
(\byear{2013})
\end{barticle}
\endbibitem

\bibitem[\protect\citeauthoryear{Nolte et~al.}{2022}]{currentapproaches}
\begin{botherref}
\oauthor{\bsnm{Nolte}, \binits{P.}},
\oauthor{\bsnm{Dullin}, \binits{C.}},
\oauthor{\bsnm{Svetlove}, \binits{A.}},
\oauthor{\bsnm{Brettmacher}, \binits{M.}},
\oauthor{\bsnm{Ru{\ss}mann}, \binits{C.}},
\oauthor{\bsnm{Schilling}, \binits{A.F.}},
\oauthor{\bsnm{Alves}, \binits{F.}},
\oauthor{\bsnm{Stock}, \binits{B.}}:
Current approaches for image fusion of histological data with computed
  tomography and magnetic resonance imaging.
Radiology Research and Practice
(2022)
\end{botherref}
\endbibitem

\bibitem[\protect\citeauthoryear{Leroy et~al.}{2023}]{Leroy}
\begin{bchapter}
\bauthor{\bsnm{Leroy}, \binits{A.}},
\bauthor{\bsnm{Cafaro}, \binits{A.}},
\bauthor{\bsnm{Gessain}, \binits{G.}},
\bauthor{\bsnm{Champagnac}, \binits{A.}},
\bauthor{\bsnm{Gr{\'e}goire}, \binits{V.}},
\bauthor{\bsnm{Deutsch}, \binits{E.}},
\bauthor{\bsnm{Lepetit}, \binits{V.}},
\bauthor{\bsnm{Paragios}, \binits{N.}}:
\bctitle{Structuregnet: Structure-guided multimodal 2d-3d registration}.
In: \beditor{\bsnm{Greenspan}, \binits{H.}},
\beditor{\bsnm{Madabhushi}, \binits{A.}},
\beditor{\bsnm{Mousavi}, \binits{P.}},
\beditor{\bsnm{Salcudean}, \binits{S.}},
\beditor{\bsnm{Duncan}, \binits{J.}},
\beditor{\bsnm{Syeda-Mahmood}, \binits{T.}},
\beditor{\bsnm{Taylor}, \binits{R.}} (eds.)
\bbtitle{Medical Image Computing and Computer Assisted Intervention -- MICCAI
  2023},
pp. \bfpage{771}--\blpage{780}.
\bpublisher{Springer},
\blocation{Cham}
(\byear{2023})
\end{bchapter}
\endbibitem

\bibitem[\protect\citeauthoryear{Chicherova et~al.}{2014}]{feature1}
\begin{bchapter}
\bauthor{\bsnm{Chicherova}, \binits{N.}},
\bauthor{\bsnm{Fundana}, \binits{K.}},
\bauthor{\bsnm{M{\"u}ller}, \binits{B.}},
\bauthor{\bsnm{Cattin}, \binits{P.C.}}:
\bctitle{Histology to $\mu$ct data matching using landmarks and a density
  biased ransac}.
In: \beditor{\bsnm{Golland}, \binits{P.}},
\beditor{\bsnm{Hata}, \binits{N.}},
\beditor{\bsnm{Barillot}, \binits{C.}},
\beditor{\bsnm{Hornegger}, \binits{J.}},
\beditor{\bsnm{Howe}, \binits{R.}} (eds.)
\bbtitle{Medical Image Computing and Computer-Assisted Intervention -- MICCAI
  2014},
pp. \bfpage{243}--\blpage{250}.
\bpublisher{Springer},
\blocation{Cham}
(\byear{2014})
\end{bchapter}
\endbibitem

\bibitem[\protect\citeauthoryear{Chicherova et~al.}{2018}]{feature2}
\begin{barticle}
\bauthor{\bsnm{Chicherova}, \binits{N.}},
\bauthor{\bsnm{Hieber}, \binits{S.E.}},
\bauthor{\bsnm{Khimchenko}, \binits{A.}},
\bauthor{\bsnm{Bikis}, \binits{C.}},
\bauthor{\bsnm{Müller}, \binits{B.}},
\bauthor{\bsnm{Cattin}, \binits{P.}}:
\batitle{Automatic deformable registration of histological slides to $\mu$ct
  volume data}.
\bjtitle{Journal of Microscopy}
\bvolume{271}(\bissue{1}),
\bfpage{49}--\blpage{61}
(\byear{2018})
\end{barticle}
\endbibitem

\bibitem[\protect\citeauthoryear{Lowe}{2004}]{Lowe2004}
\begin{barticle}
\bauthor{\bsnm{Lowe}, \binits{D.G.}}:
\batitle{Distinctive image features from scale-invariant keypoints}.
\bjtitle{International Journal of Computer Vision}
\bvolume{60}(\bissue{2}),
\bfpage{91}--\blpage{110}
(\byear{2004})
\end{barticle}
\endbibitem

\bibitem[\protect\citeauthoryear{Chen}{2023}]{imxp}
\begin{botherref}
\oauthor{\bsnm{Chen}, \binits{e.a.}}:
Deformable Registration of $\mu$CT and Histology Slide.
In: Proceedings of the 6th International Symposium on Medical Applications of
  X-ray Phase-Contrast \& Photon-Counting, Technical University of Munich, July
  2023
(2023)
\end{botherref}
\endbibitem

\bibitem[\protect\citeauthoryear{Chen}{2024}]{xnpig}
\begin{botherref}
\oauthor{\bsnm{Chen}, \binits{e.a.}}:
2D-3D Deformable Image Registration of Histology Slide and Phase-contrast
  Micro-CT.
In: Proceedings of The 6th International Conference on X-ray and Neutron Phase
  Imaging with Gratings, Shenzhen Institute of Advanced Technology, Chinese
  Academy of Sciences, April 2024
(2024)
\end{botherref}
\endbibitem

\bibitem[\protect\citeauthoryear{Ronchetti et~al.}{2023}]{DISA}
\begin{bchapter}
\bauthor{\bsnm{Ronchetti}, \binits{M.}},
\bauthor{\bsnm{Wein}, \binits{W.}},
\bauthor{\bsnm{Navab}, \binits{N.}},
\bauthor{\bsnm{Zettinig}, \binits{O.}},
\bauthor{\bsnm{Prevost}, \binits{R.}}:
\bctitle{Disa: Differentiable similarity approximation for universal multimodal
  registration}.
In: \beditor{\bsnm{Greenspan}, \binits{H.}},
\beditor{\bsnm{Madabhushi}, \binits{A.}},
\beditor{\bsnm{Mousavi}, \binits{P.}},
\beditor{\bsnm{Salcudean}, \binits{S.}},
\beditor{\bsnm{Duncan}, \binits{J.}},
\beditor{\bsnm{Syeda-Mahmood}, \binits{T.}},
\beditor{\bsnm{Taylor}, \binits{R.}} (eds.)
\bbtitle{Medical Image Computing and Computer Assisted Intervention -- MICCAI
  2023},
pp. \bfpage{761}--\blpage{770}.
\bpublisher{Springer},
\blocation{Cham}
(\byear{2023})
\end{bchapter}
\endbibitem

\bibitem[\protect\citeauthoryear{Wein et~al.}{2008}]{LC2}
\begin{barticle}
\bauthor{\bsnm{Wein}, \binits{W.}},
\bauthor{\bsnm{Brunke}, \binits{S.}},
\bauthor{\bsnm{Khamene}, \binits{A.}},
\bauthor{\bsnm{Callstrom}, \binits{M.R.}},
\bauthor{\bsnm{Navab}, \binits{N.}}:
\batitle{Automatic ct-ultrasound registration for diagnostic imaging and
  image-guided intervention}.
\bjtitle{Medical Image Analysis}
\bvolume{12}(\bissue{5}),
\bfpage{577}--\blpage{585}
(\byear{2008}).
\bcomment{Special issue on the 10th international conference on medical imaging
  and computer assisted intervention - MICCAI 2007}
\end{barticle}
\endbibitem

\bibitem[\protect\citeauthoryear{Wein et~al.}{2013}]{LC22}
\begin{botherref}
\oauthor{\bsnm{Wein}, \binits{W.}},
\oauthor{\bsnm{Ladikos}, \binits{A.}},
\oauthor{\bsnm{Fuerst}, \binits{B.}},
\oauthor{\bsnm{Shah}, \binits{A.}},
\oauthor{\bsnm{Sharma}, \binits{K.}},
\oauthor{\bsnm{Navab}, \binits{N.}}:
Global registration of ultrasound to mri using the lc2 metric for enabling
  neurosurgical guidance
\textbf{16}(Pt 1),
34--41
(2013)
\end{botherref}
\endbibitem

\bibitem[\protect\citeauthoryear{Rueckert et~al.}{1999}]{FFD}
\begin{barticle}
\bauthor{\bsnm{Rueckert}, \binits{D.}},
\bauthor{\bsnm{Sonoda}, \binits{L.I.}},
\bauthor{\bsnm{Hayes}, \binits{C.}},
\bauthor{\bsnm{Hill}, \binits{D.L.G.}},
\bauthor{\bsnm{Leach}, \binits{M.O.}},
\bauthor{\bsnm{Hawkes}, \binits{D.J.}}:
\batitle{Nonrigid registration using free-form deformations: application to
  breast mr images}.
\bjtitle{IEEE Transactions on Medical Imaging}
\bvolume{18}(\bissue{8}),
\bfpage{712}--\blpage{721}
(\byear{1999})
\end{barticle}
\endbibitem

\bibitem[\protect\citeauthoryear{Maas}{2013}]{LeakyReLU}
\begin{bchapter}
\bauthor{\bsnm{Maas}, \binits{A.L.}}:
\bctitle{Rectifier nonlinearities improve neural network acoustic models}.
In: \bbtitle{Proceedings of the International Conference on Machine Learning
  (ICML)}
(\byear{2013})
\end{bchapter}
\endbibitem

\bibitem[\protect\citeauthoryear{He et~al.}{2016}]{7780459}
\begin{bchapter}
\bauthor{\bsnm{He}, \binits{K.}},
\bauthor{\bsnm{Zhang}, \binits{X.}},
\bauthor{\bsnm{Ren}, \binits{S.}},
\bauthor{\bsnm{Sun}, \binits{J.}}:
\bctitle{Deep residual learning for image recognition}.
In: \bbtitle{2016 IEEE Conference on Computer Vision and Pattern Recognition
  (CVPR)},
pp. \bfpage{770}--\blpage{778}
(\byear{2016})
\end{bchapter}
\endbibitem

\bibitem[\protect\citeauthoryear{Zhang}{2019}]{blurpool}
\begin{bchapter}
\bauthor{\bsnm{Zhang}, \binits{R.}}:
\bctitle{Making convolutional networks shift-invariant again}.
In: \beditor{\bsnm{Chaudhuri}, \binits{K.}},
\beditor{\bsnm{Salakhutdinov}, \binits{R.}} (eds.)
\bbtitle{Proceedings of the 36th International Conference on Machine Learning}.
\bsertitle{Proceedings of Machine Learning Research},
vol. \bseriesno{97},
pp. \bfpage{7324}--\blpage{7334}
(\byear{2019})
\end{bchapter}
\endbibitem

\bibitem[\protect\citeauthoryear{}{}]{exciscope}
\begin{botherref}
Exciscope.
\url{https://exciscope.com/}
\end{botherref}
\endbibitem

\bibitem[\protect\citeauthoryear{Fletcher}{2000}]{bfgs}
\begin{bchapter}
\bauthor{\bsnm{Fletcher}, \binits{R.}}:
\bctitle{Newton-like methods}.
In: \beditor{\bsnm{Alex}, \binits{A.}} (ed.)
\bbtitle{Practical Methods of Optimization},
pp. \bfpage{44}--\blpage{79}.
\bpublisher{John Wiley \& Sons, Ltd},
\blocation{The Atrium, Southern Gate Chichester, England PO19 8SQ}
(\byear{2000}).
\bcomment{Chap. 3}
\end{bchapter}
\endbibitem

\bibitem[\protect\citeauthoryear{Powell}{2009}]{BOBYQA}
\begin{botherref}
\oauthor{\bsnm{Powell}, \binits{M.J.D.}}:
The {BOBYQA} algorithm for bound constrained optimization without derivatives.
Technical Report NA2009/06,
Department of Applied Mathematics and Theoretical Physics, Cambridge
  University,
Cambridge, UK
(2009)
\end{botherref}
\endbibitem

\bibitem[\protect\citeauthoryear{Johnson}{2007}]{NLopt}
\begin{botherref}
\oauthor{\bsnm{Johnson}, \binits{S.G.}}:
The {NLopt} nonlinear-optimization package.
\url{https://github.com/stevengj/nlopt}
(2007)
\end{botherref}
\endbibitem

\bibitem[\protect\citeauthoryear{}{}]{psi}
\begin{botherref}
TOMCAT, Swiss Light Source, Paul Scherrer Institute.
\url{https://www.psi.ch/en/sls/tomcat}
\end{botherref}
\endbibitem

\bibitem[\protect\citeauthoryear{}{}]{mitos}
\begin{botherref}
MITOS GmbH.
\url{https://www.mitos.de/en/services.html}
\end{botherref}
\endbibitem

\end{thebibliography}

\end{document}